\documentclass[sigconf]{acmart}

\usepackage{balance}
\usepackage{algorithm}
\usepackage{algorithmic}
\usepackage{graphicx}
\usepackage{pgf}
\usepackage{tikz}
\usepackage{booktabs}
\usetikzlibrary{arrows, automata, shapes,snakes}
\usepackage[latin1]{inputenc}
\usepackage{array}

\def\BibTeX{{\rm B\kern-.05em{\sc i\kern-.025em b}\kern-.08emT\kern-.1667em\lower.7ex\hbox{E}\kern-.125emX}}
    
\setcopyright{acmlicensed}

\begin{document}

\copyrightyear{2019} 
\acmYear{2019} 
\acmConference[MM '19]{Proceedings of the 27th ACM International Conference on Multimedia}{October 21--25, 2019}{Nice, France}
\acmBooktitle{Proceedings of the 27th ACM International Conference on Multimedia (MM '19), October 21--25, 2019, Nice, France}
\acmPrice{15.00}
\acmDOI{10.1145/3343031.3350874}
\acmISBN{978-1-4503-6889-6/19/10}

\fancyhead{}

%
\title{AdaCompress: Adaptive Compression for \\ Online Computer Vision Services}

%
 \author{Hongshan Li}
 
 \orcid{0000-0003-2908-9873}
 \affiliation{%
   \institution{Tsinghua-Berkeley Shenzhen Institute, Tsinghua University}
 }
\email{lhs17@mails.tsinghua.edu.cn}

\author{Yu Guo}
\affiliation{%
	\institution{Graduate School at Shenzhen, Tsinghua University}
}
\email{guoy18@mails.tsinghua.edu.cn}

\author{Zhi Wang}
\authornote{corresponding authors}
\affiliation{%
	\institution{Graduate School at Shenzhen, Tsinghua University}
	\institution{Peng Cheng Laboratory}
}
\email{wangzhi@sz.tsinghua.edu.cn}

\author{Shutao Xia}
\affiliation{%
	\institution{Graduate School at Shenzhen, Tsinghua University}
}
\email{xiast@sz.tsinghua.edu.cn}

\author{Wenwu Zhu}
\authornotemark[1]
\affiliation{%
	\institution{Tsinghua-Berkeley Shenzhen institute, Department of Computer Science and Technology, Tsinghua University}
}
\email{wwzhu@tsinghua.edu.cn}

%
 \renewcommand{\shortauthors}{Hongshan Li and Yu Guo, et al.}

%
\begin{abstract}
With the growth of computer vision based applications and services, an explosive amount of images have been uploaded to cloud servers which host such computer vision algorithms, usually in the form of deep learning models. JPEG has been used as the {\em de facto} compression and encapsulation method before one uploads the images, due to its wide adaptation. However, standard JPEG configuration does not always perform well for compressing images that are to be processed by a deep learning model, e.g., the standard quality level of JPEG leads to 50\% of size overhead (compared with the best quality level selection) on ImageNet under the same inference accuracy in popular computer vision models including InceptionNet, ResNet, etc. Knowing this, designing a better JPEG configuration for online computer vision services is still extremely challenging: 1) Cloud-based computer vision models are usually a black box to end-users; thus it is difficult to design JPEG configuration without knowing their model structures. 2) JPEG configuration has to change when different users use it. In this paper, we propose a reinforcement learning based JPEG configuration framework. In particular, we design an agent that adaptively chooses the compression level according to the input image's features and backend deep learning models. Then we train the agent in a reinforcement learning way to adapt it for different deep learning cloud services that act as the {\em interactive training environment} and feeding a reward with comprehensive consideration of accuracy and data size. In our real-world evaluation on Amazon Rekognition, Face++ and Baidu Vision, our approach can reduce the size of images by 1/2 -- 1/3 while the overall classification accuracy only decreases slightly.
\end{abstract}

%
%



%
\maketitle

\section{Introduction}
\label{sec: introduction}

With the great success of deep learning in computer vision, this decade has witnessed an explosion of deep learning based computer vision applications. Because of the huge computational resource consumption for deep learning applications (e.g., inferring an image on VGG19~\cite{VGG19} requires 20 GFLOPS GPU resource), in today's computer vision applications, users usually have to upload the input images to the central cloud service providers (e.g., SenseTime, Baidu Vision and Google Vision, etc.), leading to a significant uploading traffic burden. For example, a picture taken by a cellphone at the resolution of $3968 \times 2976$ when saved as JPEG format at the default compression level, has a size up to 3MB. 

To reduce the upload traffic, it is straightforward that an image should be compressed before one uploads it. Though JPEG has been used as the de facto image compression and encapsulation method, its performance for the deep computer vision models is not satisfactory, because JPEG was originally designed for human vision system. Liu et al.~\cite{DeepN-JPEG} showed that by modifying the \emph{quality level} in the default JPEG configuration, by retraining it on the original dataset, one can compress an image to a smaller version while maintaining the inference accuracy for a fixed deep computer vision algorithm. We then raise an intuitive question: to make it practically useful, can we improve the JPEG configuration adaptively for different cloud computer vision services, without any pre-knowledge of the original model and dataset? 

Our answer to this question is a new learning-based compression methodology for today's cloud computer vision services. We tackle the following challenges in our design. 

\begin{figure}
	\begin{minipage}{0.45\linewidth}
		\centerline{\includegraphics[width=3.0cm, trim=0 180 0 150, clip]{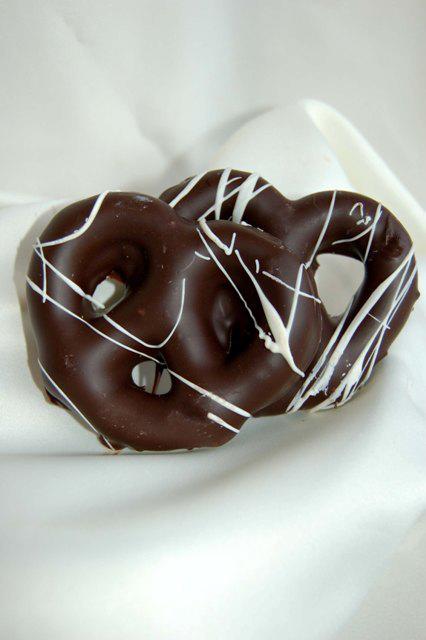}}
		\centerline{(1a) Q=75}
		\centerline{Face++ prediction \ = \ ["donut"]}
	\end{minipage}
	\hfill
	\begin{minipage}{0.45\linewidth}
		\centerline{\includegraphics[width=3.0cm, trim=0 180 0 150, clip]{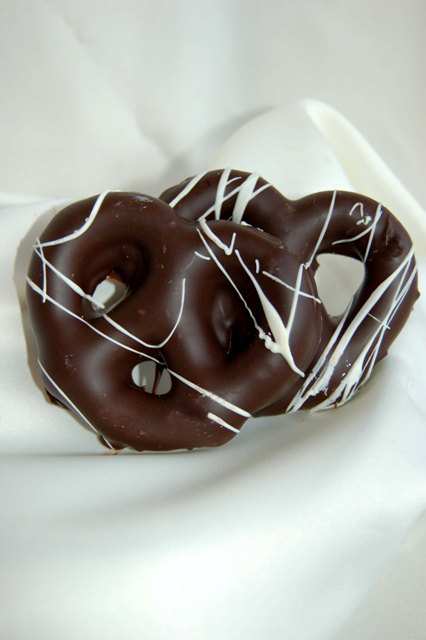}}
		\centerline{(1b) Q=55}
		\centerline{Face++ prediction \ = \ []}
	\end{minipage}
	\vfill
	\begin{minipage}{0.45\linewidth}
		\centerline{\includegraphics[width=3.0cm, trim=0 0 0 10]{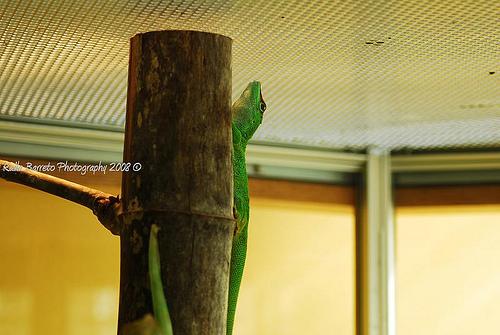}}
		\centerline{(2a) Q=75}
		\centerline{Baidu prediction \ = \ ["chameleon"]}
	\end{minipage}
	\hfill
	\begin{minipage}{0.45\linewidth}
		\centerline{\includegraphics[width=3.0cm, trim=0 0 0 10]{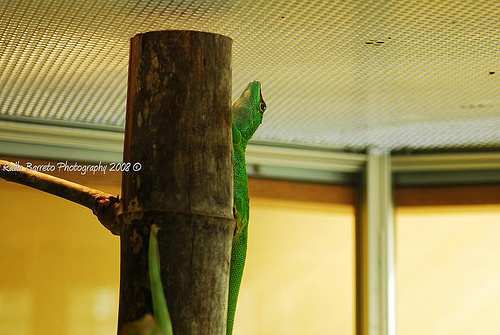}}
		\centerline{(2b) Q=55}
		\centerline{Baidu prediction \ = \ ["electric fan"]}
	\end{minipage}
	\vfill
	\begin{minipage}{0.45\linewidth}
		\centerline{\includegraphics[width=3.0cm, trim=0 20 0 20, clip]{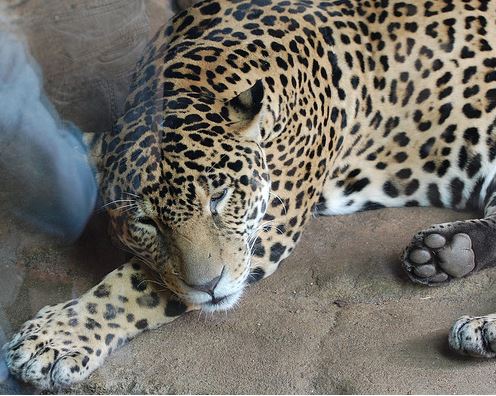}}
		\centerline{(3a) Q=75}
		\centerline{Baidu prediction \ = \ ["leopard"]}
	\end{minipage}
	\hfill
	\begin{minipage}{0.45\linewidth}
		\centerline{\includegraphics[width=3.0cm, trim=0 20 0 20, clip]{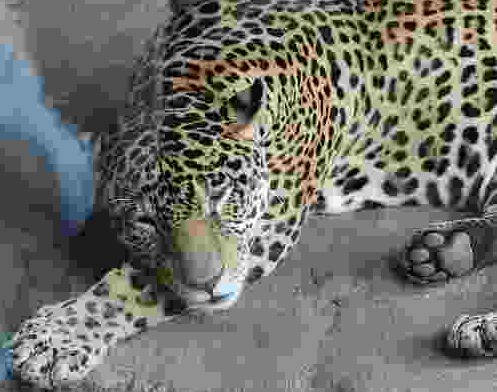}}
		\centerline{(3b) Q=5}
		\centerline{Baidu prediction \ = \ ["leopard"]}
	\end{minipage}
	\caption{The prediction of a deep learning model is not completely related to the input image's quality, making it difficult to use a fixed compression quality for all images. For image 1a, 1b and 2a, 2b, minor changes cause different predictions though they are visually similar; for image 3a and 3b, the cloud model still output correct label from a severely compressed image though they look very different}
	\label{fig: compress_accuracy}
	\vspace{-0.6cm}
\end{figure}

\begin{itemize}
\item \textbf{Lack of information about the cloud computer vision models.} Different from the studies ~\cite{DeepN-JPEG,torfason2018towards,gueguen2018faster}, in which the computer vision models are available so that one can adjust the JPEG configuration according to the model structure or retrain the parameters in it, e.g., one can greedily search a gradient descent to reach an optimal compression level in JPEG. In our study, however, the details of the online cloud computer vision model are inaccessible. 

\item \textbf{Different cloud computer vision models need different JPEG configurations.} As an adaptive JPEG configuration solution, we target to provide a solution that is adaptive to different cloud computer vision services, i.e., it can \emph{generate} JPEG configuration for different models. However, today's cloud computer vision algorithms, based one deep and convolutional computations, are quite hard to understand. The same compression level could lead to totally different accuracy performance. Some examples are shown in Figure \ref{fig: compress_accuracy}, picture 1a and 1b, 2a and 2b are visually similar for human beings, but the deep learning models give different inference results, only because they are compressed at different quality levels. And such relationship is not apparent, e.g., picture 3b is highly compressed and looks destroyed comparing to picture 3a, but the deep learning model can still recognize it. This phenomenon is also presented in ~\cite{delac2005effects} and commonly seen in adversarial neural network researches~\cite{yuan2019adversarial, evtimov2018robust}. 

\item \textbf{Lack of well-labeled training data.} In our problem, one is not provided the well-labeled data on which image should be compressed to which quality level, as in conventional supervised deep learning tasks. In practice, such an image compression module is usually utilized in an online manner, and the solution has to learn from the images it uploads automatically. 
\end{itemize} 

To address the above challenges, we present a deep reinforcement learning (DRL) based solution, AdaCompress, to choose the proper compression level for an image to a computer vision model on the cloud, in an online manner. We open-sourced\footnote{\url{https://github.com/hosea1008/AdaCompress}} our JPEG configuration module that works with today's cloud computer vision APIs upon acceptance of this paper. In particular, our contributions are summarized as follows: 

\begin{itemize} \item First, we design an interactive training environment that can be applied to different computer vision cloud services at different times, then we propose a deep Q neural network agent to evaluate and predict the performance of a compression level on an input image. In real-world application scenarios, this deep Q neural network can be highly efficient to run on today's edge infrastructures (e.g., Google edge TPU \cite{google-tpu}, Huawei Atlas 500 edge station \cite{huawei-atlas500}).
	
	\item Second, we build up a reinforcement learning framework to train the deep Q network in the above environment. By feeding the agent with carefully designed reward comprehensively considering accuracy and data size, the agent can learn to choose a proper compression level for an input image after iteratively interacting with the environment. To make the solution adaptive to the changing input images, we propose an explore-exploit mechanism to adapt the agent to different ``scenery'' online. After deploying the deep Q agent, an inference-estimate-retrain mechanism is designed to restart the training procedure once the scenery changes, and the existing running Q agent cannot guarantee stable accuracy performance. 
	
	\item Finally, we provide analysis and insights on our design. We analyze the Q network's behavior by introducing Grad-Cam ~\cite{grad-cam}, and we explain why the Q network chooses a specific compression level, and provide some general patterns. Generally speaking, images that contain large smooth areas are more sensitive to compression, while the images with complex textures are more robust to compression when shown to deep learning models. We evaluate our system on some most popular cloud deep learning services, including Amazon Rekognition~\cite{amazon_rekognition}, Face++~\cite{face++_service} and Baidu Vision~\cite{baidu_vision}, and show that our design can reduce the uplink traffic load by up to 1/2 while maintaining comparable overall accuracy. \end{itemize}

The rest of this paper is organized as follows. We present our framework and detailed design in Sec. \ref{sec: design}. In Sec. \ref{sec: evaluation} we present our solution's performance. We discuss related works in Sec. \ref{sec: related_works} and conclude the paper in Sec. \ref{sec: conclusion}.

\section{Detailed Design}
\label{sec: design}

\begin{figure}
	\begin{minipage}{\linewidth}
		\centerline{\includegraphics[width=\linewidth]{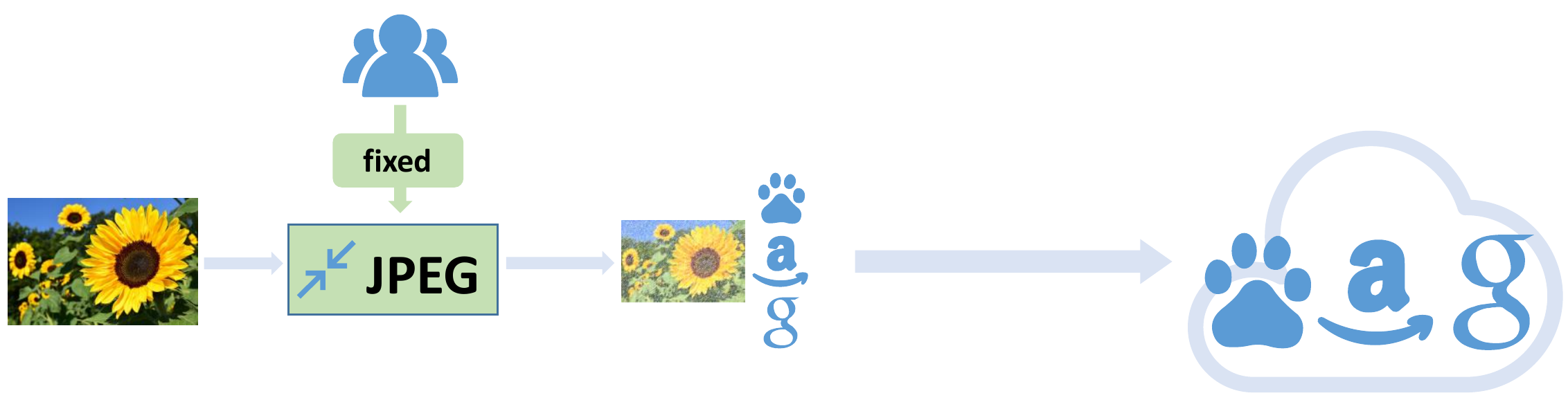}}
		\begin{center}
			{(a) Conventional solution: \\ fixed user-defined compression level}
		\end{center}
		\vspace{0.3cm}
	\end{minipage}
	
	\begin{minipage}{\linewidth}
		\centerline{\includegraphics[width=\linewidth]{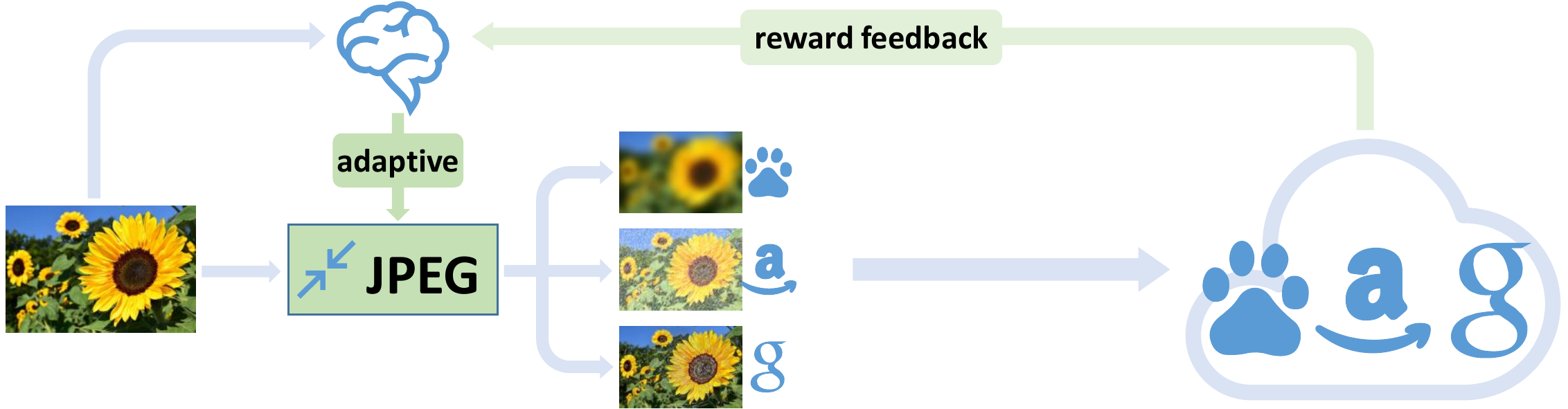}}
		\begin{center}
			{(b) AdaCompress solution: \\ input image and model aware compression}
		\end{center}
	\end{minipage}
	\caption{Comparing to the conventional solution, our solution can update the compression strategy based on the backend model feedback}
	\label{fig: framework}
\end{figure}

A brief framework of AdaCompress is shown in Figure \ref{fig: framework}. Briefly, it is a DRL (deep reinforcement learning) based system to train an agent to choose the proper quality level $ c $ for one image to be compressed by JPEG. We will discuss the formulation, agent design, reinforcement learning framework, reward feedback, and retrain mechanism separately in the following subsections. We will provide experimental details of all the hyperparameters in Sec. \ref{sec: evaluation}. 

\subsection{Problem formulation}
\label{subsec: formulation}


Without loss of generality, we denote the cloud deep learning service as $ \vec{y}_i = M(x_i) $ that provides a predicted result list $ \vec{y}_i $ for each input image $ x_i $, and it has a baseline output $ \vec{y}_{\rm ref} = M(x_{\rm ref}) $ for all reference input $ x \in X_{\rm ref} $. We use this $ \vec{y}_{\rm ref} $ as the ground truth labels, and for each image $ x_c $ compressed at quality $ c $, we have $ \vec{y}_c = M(x_c) $. Therefore, we have an accuracy metric $ \mathcal{A}_c $ by comparing $ \vec{y}_{\rm ref} $ and $ \vec{y}_c $. To be general, we use the top-5 accuracy as the following $ \mathcal{A} $, the same as the classification metric of ILSVRC2012~\cite{ILSVRC12}.

\begin{align*}
	\mathcal{A} =& \sum_{k}\min_jd(l_j, g_k) \\
	& l_j \in \vec{y}_c, \quad j=1,...,5 \\
	& g_k \in \vec{y}_{\rm ref}, \quad k=1, ..., {\rm length}(\vec{y}_{\rm ref}) \\
	& d(x, y) = 1 \ \text{if} \ x=y  \ \text{else} \ 0 
\end{align*}

Where $ j = 1,...,5 $ indicating the prediction labels at top-5 score, $ k = 1,...,{\rm length}(\vec{y}_{\rm ref}) $ means that if anyone of the top-5 predicted labels matches one of the predictions from $ \vec{y}_{\rm ref} $, it is regarded as a correct prediction. To be general, we stipulate that for a cloud deep learning service, we cannot get the deep model's in-layer details (e.g., softmax probabilities) therefore we use a binary hard label $ d(x, y) \in \{0, 1\} $ to evaluate the accuracy.

We also denote JPEG input images as $ f_{ic} = J(x_i, c) $ that for an input image $ x_i $ and a given compression quality $ c $, it outputs a compressed file $ f_{ic} $ at the size of $ s_{ic} $, for a reference compression level $ c_{\rm ref} $, the compressed file size is $ s_{\rm ref} $. Besides, images input from a specific location usually belong to a particular contextual group. For example, in an indoor scenery, the user input is less likely to have the images of the ocean, airplanes, and dolphins but more likely to have furniture and so on. Therefore, the agent at one place does not need to know all the contextual features in all places. We formulated this as contextual group $ \mathcal{X} $. This contextual grouping concept is also discussed in ~\cite{mcdnn}.

Initially, the agent tries different compression level $ c_{\min} < c < c_{\max}, c \in \mathbb{N} $ to obtain compressed image $ x_c $ from input image $ x $, and an image compressed at a reference level $ c_{\rm ref} $ is also uploaded to the cloud to obtain $ \vec{y}_{\rm ref} $. Comparing the two uploaded instances $ \{x, x_c\} $ and cloud recognition results $ \{\vec{y}_{\rm ref}, \vec{y}_c\} $, we can have the reference file size $ s_{\rm ref} $ and compressed file size $ s_c $ and therefore the file compression ratio $ \Delta s = \frac{s_c}{s_{\rm ref}} $ and accuracy metric $ \mathcal{A}_c $.

\subsection{DRL agent design}

The DRL agent is expected to give a proper compression level $ c $ that minimizing the file size $ s_c $ while keeping the accuracy $ \mathcal{A} $. For the DRL agent, the input features are continuous numerical vectors, and the expected output are discrete quality levels $ c $, therefore we can use DQN (Deep Q Network)~\cite{DQN} as the DRL agent. But naive DQN can't work well in this task because of the following challenges: 

\begin{itemize}
	\item The state space of reinforcement learning is too large, and to preserve enough details, we have to add many layers and nodes to the neural network, making the DRL agent extremely difficult to converge. 
	\item It takes a long time to train one step in a large inference neural network, making the training process too time-consuming.
	\item DRL starts training from random trials, and starts learning after it found a better reward feedback. When training from a randomly initialized neural network, the reward feedback is very sparse, making it difficult for the agent to learn.
\end{itemize}

To address these challenges, we use the early layers of a well-trained neural network to extract the structural information of an input image. This is a commonly used strategy in training a deep neural network~\cite{finetunning,finetunning2}. Therefore instead of training a DRL agent directly from the input image, we use a pre-trained small neural network to extract the features from the input image to reduce the input dimension and accelerate the training procedure. In this work, we use the early convolution layers of MobileNetV2~\cite{MobileNetV2} as the image feature extractor $ \mathcal{E}(\cdot) $ for its efficiency in image classification and lightweight. The Q network $ \phi $ is connected to the feature extractor's last convolution layer, therefore the output of $ \mathcal{E} $ is the input of $ \phi $. We update the RL agent's policy by changing the parameters of Q network $ \phi $ while the feature extractor $ \mathcal{E} $ remains fixed. 

\subsection{Reinforcement learning framework}

In a specific scenery where the user input $ x $ belongs context group $ \mathcal{X} $, we define the contextual information $ \mathcal{X} $, along with the backend cloud model $ M $, as the \emph{emulator environment} $ \{\mathcal{X}, M\} $ of the reinforcement learning problem. 

Based on this insight, we formulate the feature extractor's output $ \mathcal{E}(J(\mathcal{X}, c)) $ as \emph{states}, and the compression quality $ c $ as discrete \emph{actions}. In our system, to accelerate training, we define 10 discrete actions to indicate 10 quality levels of JPEG ranging from $ 5, 15, ...,95 $. We denote the \emph{action-value function} as $ Q(\phi(\mathcal{E}(f_t)), c; \theta) $, then the optimal compression level at time $ t $ is $ c_t = {\rm argmax}_cQ(\phi(\mathcal{E}(f_t)), c; \theta) $ where $ \theta $ indicates the parameters of Q network $ \phi $. In such reinforcement learning formulation, the training phase is to minimize a loss function $ L_i(\theta_i) = \mathbb{E}_{s, c \sim \rho (\cdot)}\Big[\big(y_i - Q(s, c; \theta_i)\big)^2 \Big] $ that changes at each iteration $ i $ where $ s = \mathcal{E}(f_t) $, and $ y_i = \mathbb{E}_{s' \sim \{\mathcal{X}, M\}} \big[ r + \gamma \max_{c'} Q(s', c'; \theta_{i-1}) \mid s, c \big] $ is the target for iteration $ i $, $ r $ is the feedback reward and $ \rho(s, c) $ is a probability distribution over sequences $ s $ and quality level $ c $~\cite{DQN}. When minimizing the distance of the action-value function's output $ Q(\cdot) $ and target $ y_i $, the action-value function $ Q(\cdot) $ outputs a more accurate estimation of an action. In such formulation, it is similar to DQN problem but not the same. Different from conventional reinforcement learning, the interactions between the agent and environment are infinite; there is no signal from the environment telling that an episode has finished. Therefore, we train the RL agent intermittently at a manual interval of $ T $ after the condition $ t \geq T_{\rm start} $ guaranteeing that there are enough transitions in the memory buffer $ \mathcal{D} $. In the training phase, the RL agent firstly takes some random trials to observe the environment's reaction, and we decrease the randomness when training. All transitions are saved into a memory buffer queue $ \mathcal{D} $, the agent learns to optimize its action by minimizing the loss function $ L $ on a minibatch from $ \mathcal{D} $. The training procedure will converge as the agent's randomness keeps decaying. Finally, the agent's action is based on its historical optimal experiences. The training procedure is presented in Algorithm ~\ref{alg: rl-train}, we list the parameters in Sec.~\ref{sec: evaluation}.

\begin{algorithm}[htbp]
	\caption{Training RL agent $ \phi $ in environment $ \{\mathcal{X}, M\} $}
	\label{alg: rl-train}
	\begin{algorithmic}[1]
		\STATE Initialize replay memory queue $ \mathcal{D} $ to capacity $ N $
		\STATE Initialize action-value function $ Q $ with random weights $ \theta $
		\STATE Initialize sequence $ s_1 = \mathcal{E}\big(J(x_1, c_1)\big), x_1 \in \mathcal{X} $ and $ \phi_1 = \phi(f_1) $
		\FOR {t = 1, $ K $}
			\STATE With probability $ \epsilon $ select a random compression level $ c_t $ otherwise select $ c_t = {\rm argmax}_cQ\Big(\phi\big(\mathcal{E}(f_t)\big), c; \theta\Big) $
			\STATE Compress image $ x_t $ at quality $ c_t $ and upload it to the cloud to get result $ (\vec{y}_{\rm ref}, \vec{y}_c) $ and calculate reward $ r = R(\Delta s, \mathcal{A}_c) $
			\STATE Set $ s_{t+1} = s_t $, generate $ c_t, x_{t+1} $ and preprocess $ \phi_{t+1} = \phi \big(\mathcal{E}(f_{t+1})\big) $
			\STATE Store transition $ (\phi_t, c_t, r_t, \phi_{t+1}) $ in $ \mathcal{D} $
			\IF {$ t \mod T == 0 $ and $ t \geq T_{\rm start} $}
				\STATE Sample random minibatch of transitions $ (\phi_j, c_j, r_j, \phi_{j+1}) $ from memory buffer $ \mathcal{D} $
				\STATE Set $ y_i = r_j + \gamma \max_{c'}Q(\phi_{j+1}, c'; \theta) $
				\STATE Decay exploration rate $ \epsilon = 
					\begin{cases}
						\mu_{\rm dec}\cdot \epsilon & \text{ if } \ \mu_{\rm dec}\cdot \epsilon > \epsilon_{\min} \\ 
						\epsilon_{\min} 			& \text{ if } \ \mu_{\rm dec}\cdot \epsilon \leq \epsilon_{\min}
					\end{cases} $
				\STATE Perform a gradient descent step on $ \big(y_j - Q(\phi_j, c_j; \theta)\big)^2 $ according to ~\cite{DQN}
			\ENDIF
		\ENDFOR
	\end{algorithmic}
\end{algorithm}

\subsection{Feedback reward design}

In our solution, the agent is trained by the reward feedback from the environment $ \{\mathcal{X}, M\} $. In the above formulation, we defined compression rate $ \Delta s = \frac{s_c}{s_{\rm ref}} $ and accuracy metric $ \mathcal{A}_c $ in compression quality $ c $. Basically, we want the agent to choose a proper compression level that minimizing the file size while remaining acceptable accuracy, therefore the overall reward $ r $ should be in proportion to the accuracy $ \mathcal{A} $ while in inverse proportion to the compression ratio $ \Delta s $. We introduce two linear factors $ \alpha $ and $ \beta $ to form a linear combination $ r = \alpha \mathcal{A} - \Delta s + \beta $ as the reward function $ R(\Delta s, \mathcal{A}) $. 

\subsection{Inference-estimate-retrain mechanism}

\begin{figure}[H]
	\includegraphics[width=\linewidth]{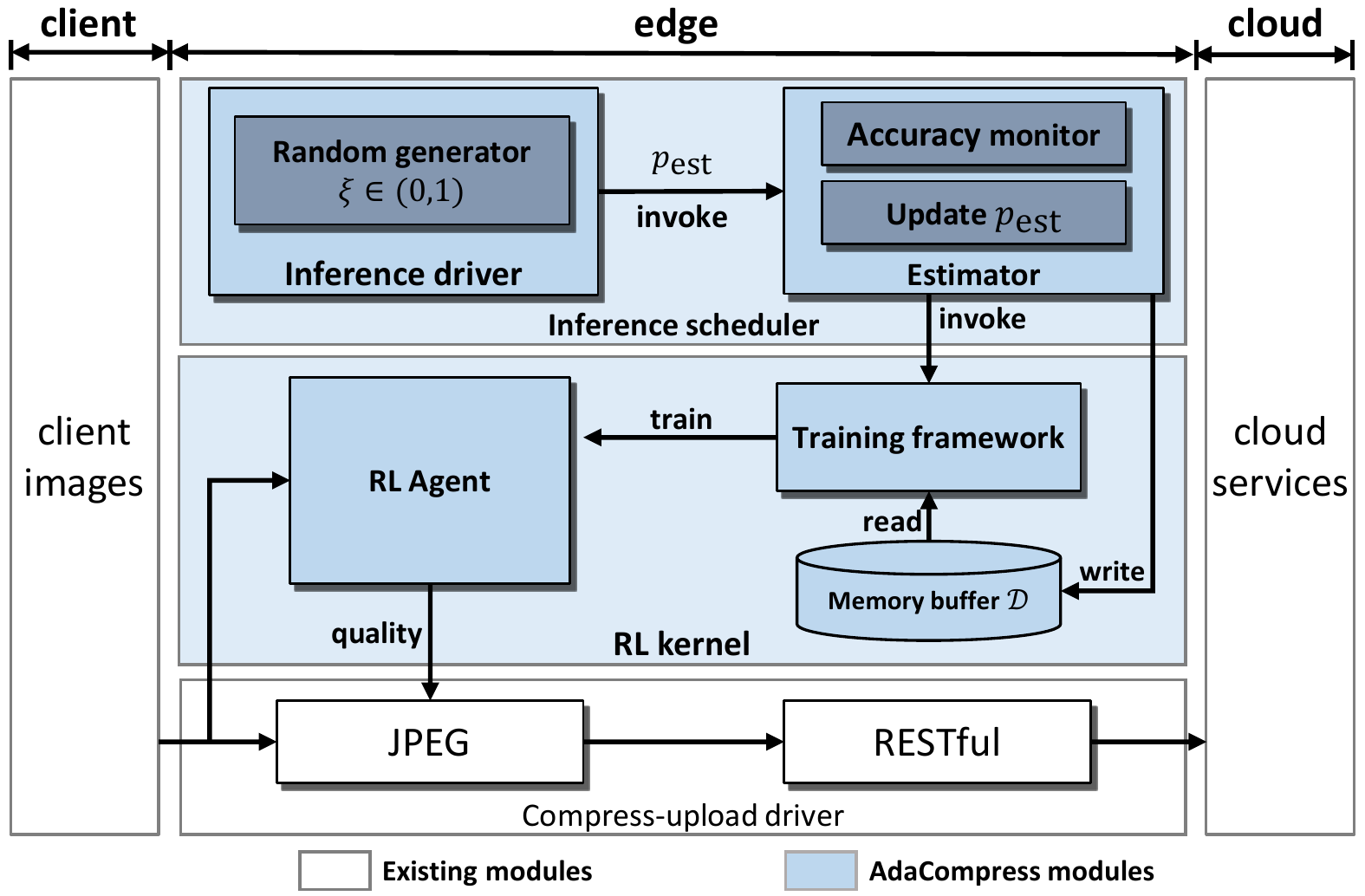}
	\caption{Diagram of AdaCompress architecture}
	\label{fig: diagram}
\end{figure}

As a running system, we introduce a running-estimate-retrain mechanism to cope with the scenery change in the inference phase, building a system with different components to inference, capturing scenery change, then retraining the RL agent. The overall system diagram is illustrated in Figure \ref{fig: diagram}.

The system diagram is shown in Figure ~\ref{fig: diagram}. We build up the memory buffer $ \mathcal{D} $ and RL (reinforcement learning) training kernel based on the compression and upload driver. When the RL kernel is called, it will load transitions from the memory buffer $ \mathcal{D} $ to train the compression level predictor $ \phi $. When the system is deployed, the pre-trained RL agent $ \phi $ guides the compression driver to compress the input image with an adaptive compression quality $ c $ according to the input image, then uploads the compressed image to cloud. 

After the AdaCompress is deployed, the input images scenery context $ \mathcal{X} $ may change. (e.g., day to night, sunny to rainy), when the scenery changes, the older RL agent's compression selection strategy may not be suitable anymore, causing the overall accuracy decreases. To cope with this scenery drifting issue, we invoke an estimator with probability $ p_{\rm est} $. We do this by generating a random value $ \xi \in (0,1) $ and compare it to $ p_{\rm est} $. If $ \xi \leq p_{\rm est} $ the estimator is invoked, AdaCompress will upload the reference image $ x_{\rm ref} $ along with the compressed image $ x_i $ to fetch $ \vec{y}_{\rm ref} $ and $ \vec{y}_i $ and therefore calculates $ \mathcal{A}_i $, and save the transition $ (\phi_i, c_i, r_i, \mathcal{A}_i) $ to the memory buffer $ \mathcal{D} $. The estimator will also compare the recent $ n $ steps' average accuracy $ \bar{\mathcal{A}}_n $ and the earliest average accuracy $ \mathcal{A}_0 $ in memory $ \mathcal{D} $, once the recent average accuracy is much lower than the initial average accuracy, the estimator will invoke the RL training kernel to retrain the agent. And once the estimator discover that the trained reward is higher than a threshold, it will stop the training kernel, returning to normal inference state. 

\begin{figure}
\begin{tikzpicture}[->,>=stealth',shorten >=1pt,auto,node distance=1.8cm, semithick]
\tikzstyle{every state}=[ellipse, align=center, draw=blue, text=black]

\node[initial, state] (B)                    {inference};
\node[state]         (C) [below right of=B] {retrain};
\node[state]         (D) [above right of=C] {estimate};

\path   (B) edge [loop above] node {$ \xi > p_{\rm est} $} (B)
		(B.10)	edge				node {$ \xi \leq p_{\rm est} $} (D.170)
		(C) edge [below left]      node {$ \bar{r}_n > r_{\rm th} $} (B)
			edge [loop below] node {$ \bar{r}_n \leq r_{\rm th} $}	 (C)
		(D) edge [loop right] node {$ \xi \leq p_{\rm est} $} (D)
		(D.190)	edge              node {$ \xi > p_{\rm est} $} (B.-10)
		(D.225)	edge [below right]      node {$ \bar{\mathcal{A}_n} < \mathcal{A}_0 $} (C);
\end{tikzpicture}
\vspace{-0.3cm}
\caption{State switching policy}
\label{fig: state-switching}
\end{figure}
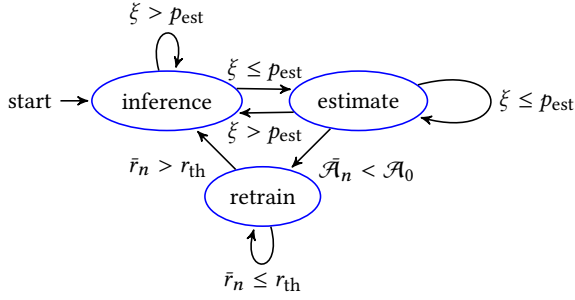

Basically, AdaCompress will adaptively switch itself between three states. The switching policy is shown as Figure~\ref{fig: state-switching}.

\subsubsection{\textbf{Inference:}}

For most times, AdaCompress runs in this state. In this state, only the compressed images are uploaded to the cloud to achieve minimum uploading traffic load. To keep a stable accuracy performance even the input scenery changes, the agent will occasionally switch to estimation state with probability $ p_{\rm est} $, meanwhile remains inference state with probability $ 1 - p_{\rm est} $. 

\subsubsection{\textbf{Estimate:}}

In this state, the reference image $ x_{\rm ref} $ and compressed image $ x_i $ are uploaded to the cloud simultaneously to fetch $ \vec{y}_{\rm ref} $ and $ \vec{y}_i $ and therefore $ \mathcal{A}_i $. In each epoch $ i $ the transition $ (\phi_i, c_i, r_i, \mathcal{A}_i) $ is logged in a memory buffer $ \mathcal{D} $. Once the average accuracy $ \bar{\mathcal{A}}_n $ of the latest $ n $ steps is lower than the average accuracy $ \mathcal{A}_0 $ of the earliest $ n $ steps in the memory buffer $ \mathcal{D} $, indicating that the current agent is no more suitable for the current input scenery, AdaCompress will switch into retrain state and invoke the RL training kernel. Otherwise, it remains estimate state with probability $ p_{\rm est} $ or switches back into inference state with probability $ 1 - p_{\rm est} $. 

Therefore, the estimating probability $ p_{\rm est} $ is vital to the whole system. On the one hand, the estimator should be invoked occasionally to estimate the current agent's accuracy, so that to retrain the agent on time once the scenery changes; on the other hand, the estimator will upload the reference image $ x_{\rm ref} $ along with the compressed image, therefore the upload size is greater than the conventional benchmark solution, causing higher traffic load. 

To trade-off between the risk of scenery changes and the objective of reducing upload traffic, we design an accuracy-aware dynamic $ p_{\rm est} $ solution, we first define that after running for $ N $ steps, the recent $ n $ steps' average accuracy is: 

\begin{align*}
	\bar{\mathcal{A}_n} &=
	\begin{cases}
		\frac{1}{n}\sum_{i=N-n}^{N} \mathcal{A}_i & \text{ if } N \geq n \\ 
		\frac{1}{n}\sum_{i=1}^{n} \mathcal{A}_i & \text{ if } N < n 
	\end{cases}
\end{align*}

With this definition, an intuitive formulation of the changes of $ p_{\rm est} $ is in inverse proportion of the gradient of $ \mathcal{A} $, meaning that when the recent accuracy is going down, we should increase the estimation probability $ p_{\rm est} $. We formulate that $ p'_{\rm est} = p_{\rm est} + \omega \nabla \bar{\mathcal{A}} $ where $ \omega $ is a scaling factor. With this recursive formula, we have the general term of $ p_{\rm est} $ with an initial estimation probability $ p_0 $ is $p_{\rm est} = p_0 + \omega \sum_{i=0}^{N} \nabla \bar{\mathcal{A}_i} $.

\subsubsection{\textbf{Retrain:}}

This state is to adapt the agent to the current input image scenery by retraining it with the memory buffer $ \mathcal{D} $, which is similar to the training procedure. The retrain phase finishes upon the recent $ n $ steps' average reward $ \bar{r}_n $ higher than a user-defined threshold $ r_{\rm th} $. And when the retrain procedure finishes, the memory buffer $ \mathcal{D} $ will be flushed, preparing to save new transitions for the retraining of a next scenery drift.

\subsection{Insight of RL agent's behavior}
\label{subsec:insight}

In the inference phase, the pre-trained RL agent predicts a proper compression level according to the input image's feature. The reference image is not uploaded to the cloud anymore; only the compressed image is uploaded, therefore, the upload traffic is reduced. We noticed that the RL agent's behavior are various for different input dataset and backend cloud services, we try to take further investigations by plotting the RL agent's "attention map" (i.e., visual explanations of why the agent chooses a quality level). 

\subsubsection{\textbf{Compression level choice variation:}}

\begin{figure}[htb]
	\includegraphics[width=\linewidth]{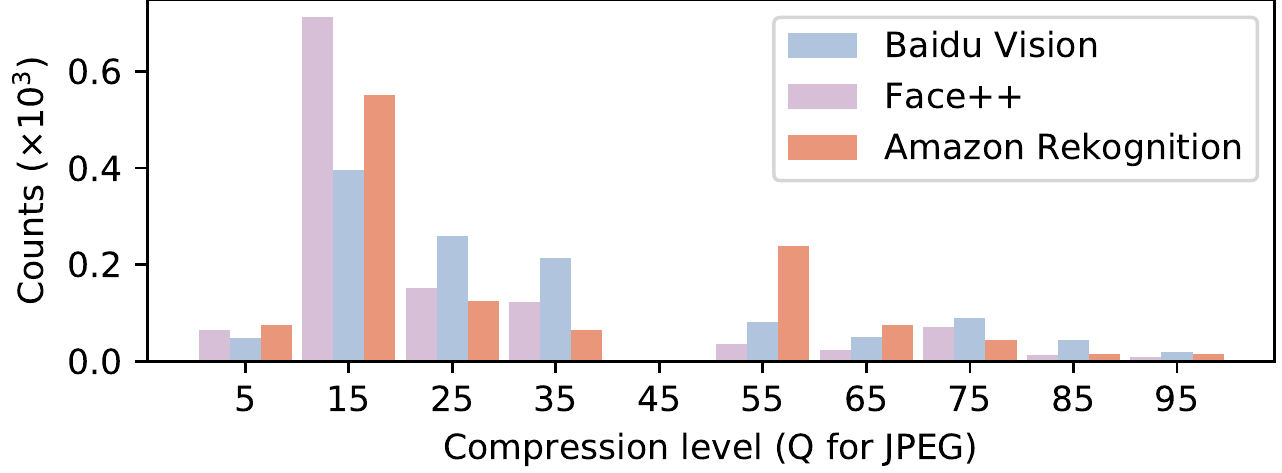}
	\caption{Histogram of RL agent's best compression level selection for different cloud services}
	\label{fig: quality_chosen}
	\vspace{-0.3cm}
\end{figure}

In our experiment, we found that in different cloud application environments, the agent's final chosen compression qualities can be quite different. As shown in Figure \ref{fig: quality_chosen}, for Face++ and Amazon Rekognition, the agent's choices are concentrated at around $ c=15 $, but for Baidu Vision, the agent's choices are distributed more evenly. Therefore, the optimal compression strategy should be different for different backend cloud services. This variation is caused by the interaction between the agent and the backend model in the training phase. Since the agent's training procedure is based on a specific backend cloud model $ M_1 $, for another cloud model $ M_2 $, the interaction between the agent and $ M_2 $ is quite different. Therefore the agent's compression level choice presents variation for different backend cloud models.  

\begin{figure}[htb]
	\includegraphics[width=\linewidth]{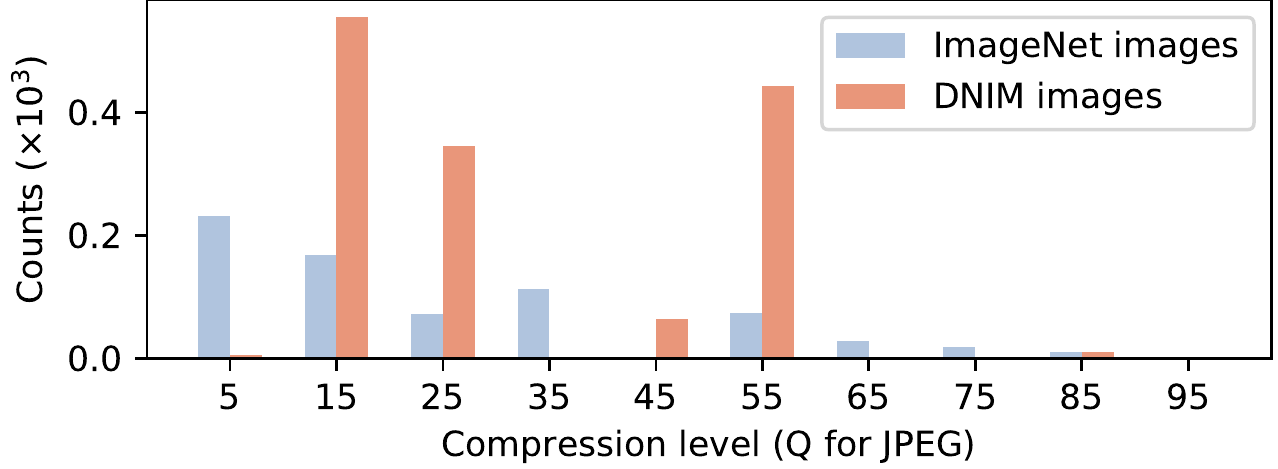}
	\caption{Histogram of RL agent's best compression level selection for different scenery image inputs}
	\label{fig: dataset_change}
\end{figure}

Moreover, in our experiment, the agent presents different behavior when the input images change from one dataset to another. Figure \ref{fig: dataset_change} shows the agent's choices for a same backend model (Baidu Vision) but different image datasets. We prepare two datasets indicating two contextual scenery. We randomly sample images from ImageNet~\cite{ImageNet} whose images are mostly taken in the daytime, to act as a daytime scenery, and we randomly select nighttime images from DNIM~\cite{DNIM} to form another dataset to act as a nighttime scenery. The histogram shown in Figure~\ref{fig: dataset_change} points out that, for the ImageNet images, the agent prefers a lower compression level, but its choices are distributed more evenly. For DNIM images, the agent's choices are more accumulated in some relatively high compression qualities. We can see that, to maintain high accuracy, when the input image's contextual group $ \mathcal{X} $ changes, the agent's compression level selection changes as well. This phenomenon presents that the agent can adaptively choose a proper compression level based on the input image's features. 

\subsubsection{\textbf{Attention map variation:}}

\begin{figure*}[htbp]
	\begin{minipage}{0.2\linewidth}
		\centerline{\includegraphics[width=3.0cm, height=2.0cm]{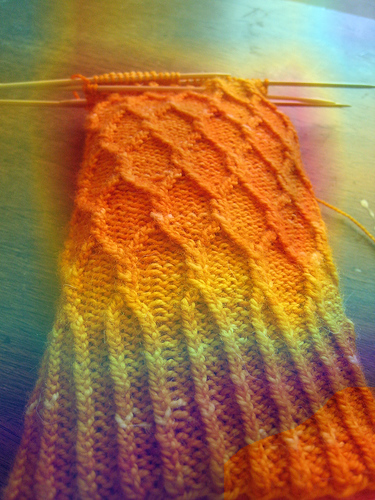}}
		\centerline{(1a) Q=5}
	\end{minipage}
	\hfill
	\begin{minipage}{0.2\linewidth}
		\centerline{\includegraphics[width=3.0cm, height=2.0cm]{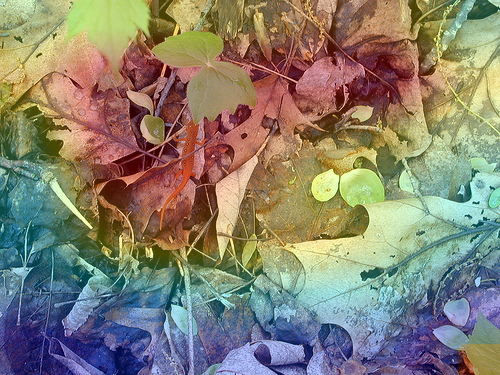}}
		\centerline{(1b) Q=15}
	\end{minipage}
	\hfill
	\begin{minipage}{0.2\linewidth}
		\centerline{\includegraphics[width=3.0cm, height=2.0cm]{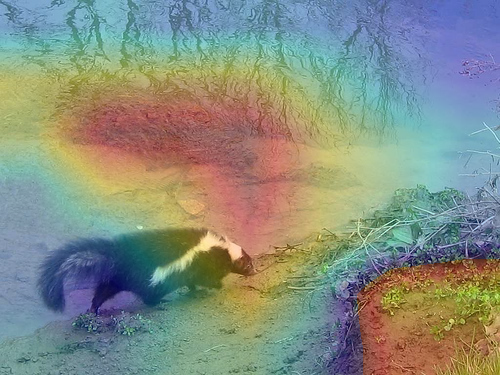}}
		\centerline{(1c) Q=15}
	\end{minipage}
	\hfill
	\begin{minipage}{0.2\linewidth}
		\centerline{\includegraphics[width=3.0cm, height=2.0cm]{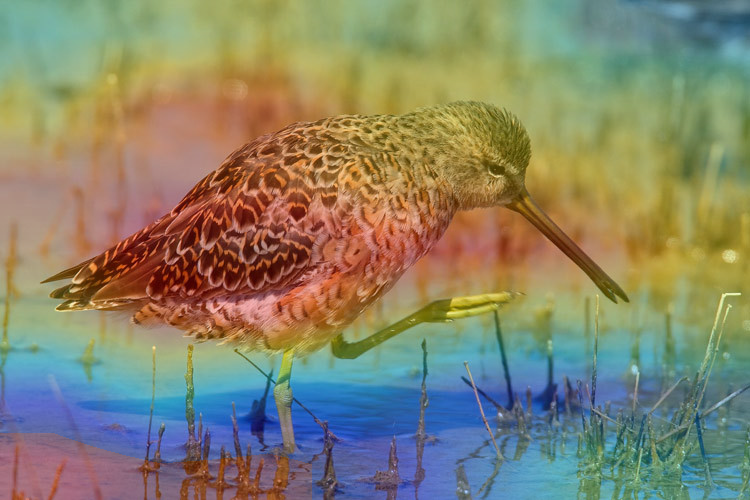}}
		\centerline{(1d) Q=15}
	\end{minipage}
	
	\vfill
	
	\begin{minipage}{0.2\linewidth}
		\centerline{\includegraphics[width=3.0cm, height=2.0cm]{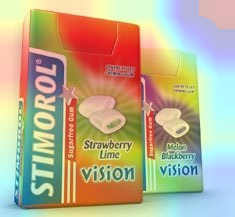}}
		\centerline{(2a) Q=85}
	\end{minipage}
	\hfill
	\begin{minipage}{0.2\linewidth}
		\centerline{\includegraphics[width=3.0cm, height=2.0cm]{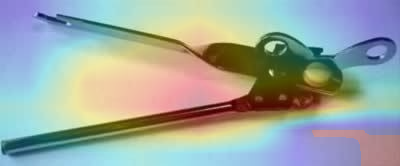}}
		\centerline{(2b) Q=85}
	\end{minipage}
	\hfill
	\begin{minipage}{0.2\linewidth}
		\centerline{\includegraphics[width=3.0cm, height=2.0cm]{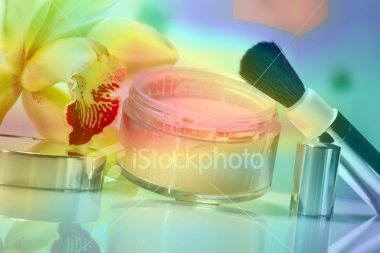}}
		\centerline{(2c) Q=75}
	\end{minipage}
	\hfill
	\begin{minipage}{0.2\linewidth}
		\centerline{\includegraphics[width=3.0cm, height=2.0cm]{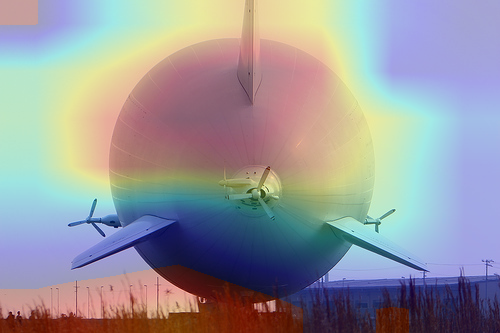}}
		\centerline{(2d) Q=75}
	\end{minipage}
	\caption{Visualization of the importance map for the RL agent to choose a compression quality}
	\label{fig: attention}
\end{figure*}

To take insight investigation, we plot the importance map of a chosen compression quality. We do so by introducing a conventional visualize algorithm, Grad-Cam, to observe the Q prediction network's interest when choosing compression levels. Grad-Cam is a widely used solution to present the importance map of a deep neural network, it is done by calculating the gradients of each target concept and backtracking to the final convolution layer. In this work, we plot the RL agent's attention map by Grad-Cam in Figure~\ref{fig: attention}. 

In our investigation, we found that in different environment $ \{\mathcal{X}, M\} $, the Q agent picks up compression qualities based on the visual textures of different regions in the image. As shown in Figure \ref{fig: attention}, picture 1a -- 1d are some pictures that the agent chooses to compress highly, the agent selects lower compression qualities based on the complex texture of the images. On the contrary, for pictures 2a -- 2d, the agent chooses higher compression qualities to preserve more details, and the agent's interest falls on some smooth regions. Especially for 1a and 2a, in picture 1a, the agent chooses a low compression level based on the rough central region though there are smooth regions around it, and in picture 2a, the agent chooses a relatively higher compression level based on the surrounding smooth region rather than the central region. 

\section{Evaluation}
\label{sec: evaluation}

In this section, we present AdaCompress's behavior and effectiveness by some real-world experiments. 

\subsection{Experiment setup}

We carry out real-world experiments to verify our solution's performance. We used a desktop PC with an NVIDIA 1080ti graphic card as the edge infrastructure. For the cloud deep learning services, we choose Baidu Vision, Face++ object detection service, and Amazon Rekognition. In the experiments, we use two datasets mentioned before in Sec.\ref{subsec:insight}, ImageNet dataset indicating daytime scenery and DNIM dataset indicating nighttime scenery. Some important hyperparameters in our experiments are given in Table \ref{tab: parameters}.

\begin{table}[]
	\begin{tabular}{llllll}
		\toprule
		notation          & value & & & notation     & value  \\ \midrule
		$c_{\rm ref}$ & 75    & & & $K$      & 1000   \\
		$\epsilon_{\min}$    & 0.02  & & & $p_0$    & 0.2    \\
		$\gamma$      & 0.95  & & & $\omega$ & -3   \\
		$ \mu_{\rm dec} $ & 0.99 & & & $ T $ & 5  \\
		$r_{\rm th}$  & 0.45   & & &   $ n  $  &  10      \\ \bottomrule
	\end{tabular}
	\caption{Experiment parameter settings}
	\label{tab: parameters}
	\vspace{-0.3cm}
\end{table}

\subsection{Metrics}

In industry, the default compression quality for JPEG is usually 75 ~\cite{pillow_benchmark,imgmin}, we regard this as a typical value $ c_{\rm ref} = 75 $ of the conventional industry benchmark. 

In our experiments, we measure the compressed and original image's file size to obtain the compression rate $ \Delta s $. Since we don't have the real ground truth label of an image, we use the output from a reference image $ \vec{y}_{\rm ref} $ as the ground truth label, and calculate the relative top-5 accuracy $ \mathcal{A} $ as the accuracy metric, the formula of $ \mathcal{A} $ is presented in Sec. ~\ref{subsec: formulation}.

\subsection{Upload size overhead}

\begin{figure}[htbp]
	\includegraphics[width=\linewidth]{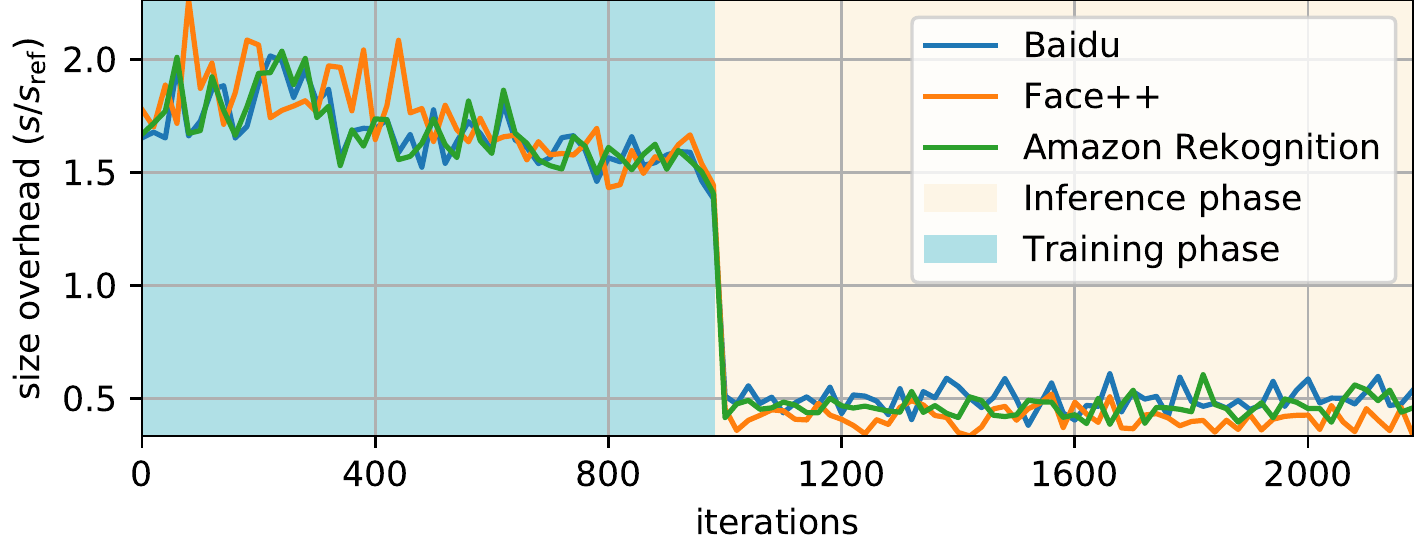}
	\caption{Size overhead in training and inference phase}
	\label{fig: train_steps}
	\vspace{-0.3cm}
\end{figure}

Figure \ref{fig: train_steps} presents the upload traffic load of the training and inference phase, to be more intuitionistic, we plot the size overhead $ \frac{s}{s_{\rm ref}} $ as the $ y $-axis where $ s $ is the real upload size of AdaCompress, $ s_{\rm ref} $ is the benchmark upload size, therefore $ y \geq 1 $ means that our solution uploads more data then benchmark, and $ y < 1 $ means the compression rate of AdaCompress. From Figure \ref{fig: train_steps} we can see that as the training procedure runs, the uploaded size is decreasing because the DRL agent is learning to choose better quality levels to upload less data. In the training phase, to train the agent while remaining a convincing recognition result, we have to upload the original image to the cloud to get the real result, along with the compressed image to obtain reward feedback, therefore the upload traffic load is even higher than the conventional solution. But once the training phase finished, the upload traffic is much lower than the benchmark. As shown in Figure \ref{fig: compress_performance}, in the inference phase, AdaCompress's upload size is only 1/2 of the benchmark's. 

\subsection{Size reduction and accuracy performance}

\begin{figure}[htbp]
	\includegraphics[width=\linewidth]{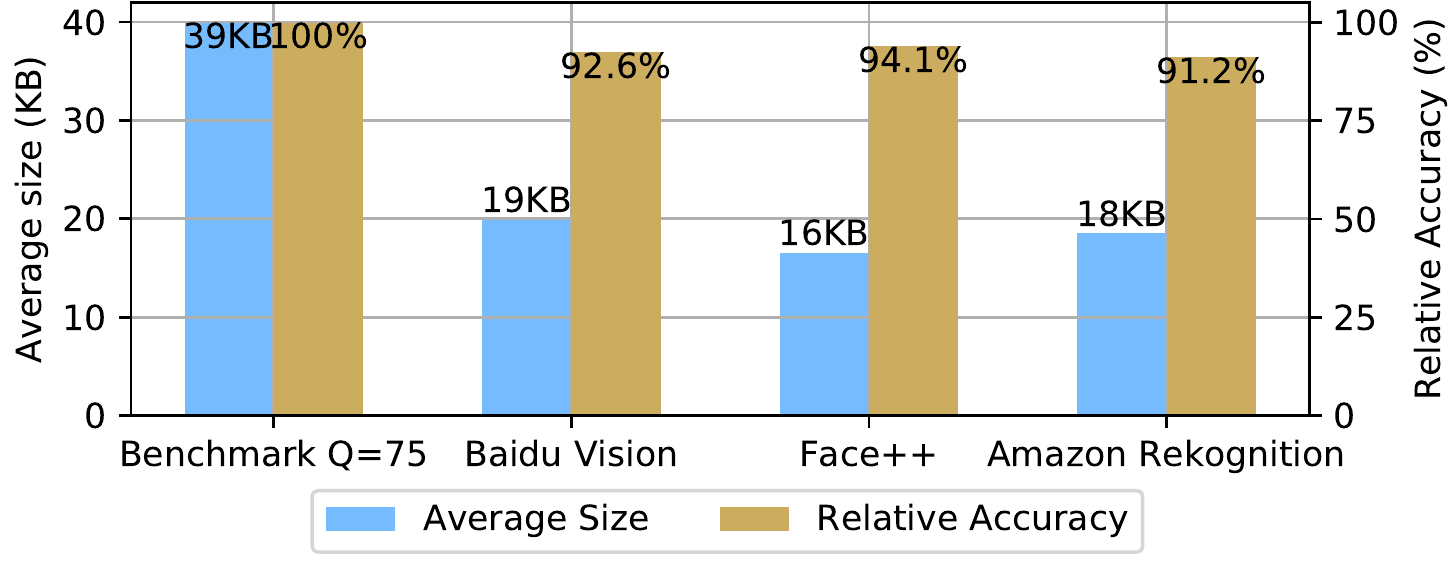}
	\caption{Average size and relative accuracy on different cloud services}
	\label{fig: compress_performance}
	\vspace{-0.3cm}
\end{figure}

Figure \ref{fig: compress_performance} presents the compression performance in the inference phase for each cloud service. We tested AdaCompress on Face++, Baidu Vision and Amazon Rekognition, comparing to the conventional compression level, for all tested cloud services, our solution can reduce the upload size by more than 1/2, meanwhile, the relative accuracy, indicated by brown bars, only decrease about 7\% on average, proving the efficiency of our design. 

\subsection{Adaptive retrain upon scenery change}

To evaluate the efficiency of the inference-estimate-retrain mechanism, we feed AdaCompress with a combined dataset whose first 720 images from DNIM night images, the later 2376 images randomly sampled from ImageNet. We adapt AdaCompress's current DRL agent to DNIM night scenery by training it on DNIM dataset, then we run AdaCompress on the combine dataset, observing AdaCompress's behavior upon the scenery change at step 720. 

\begin{figure}[htbp]
	\includegraphics[width=\linewidth]{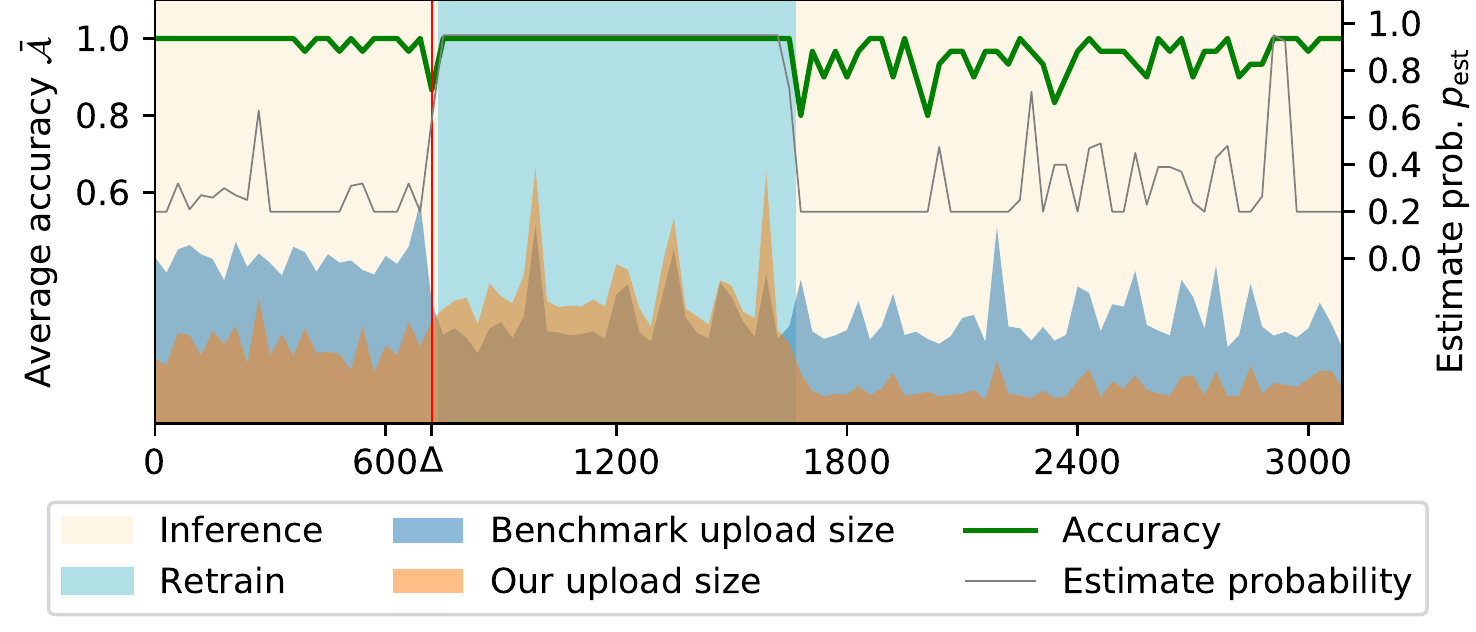}
	\caption{AdaCompress's reaction upon scenery change}
	\label{fig: running-retrain}
\end{figure}

We illustrate AdaCompress's behavior in Figure \ref{fig: running-retrain}, the $ x $-axis indicates steps, the vertical red line with a $ \Delta $ mark on $ x $-axis means the dataset change(i.e. scenery change). We plot AdaCompress's overall accuracy as the green line and the estimating probability $ p_{\rm est} $ as the gray line. At the bottom of Figure \ref{fig: running-retrain}, we also plot the scaled uploading data size of AdaCompress and benchmark solution to illustrate the upload data size overhead in the inference phase.

From Figure \ref{fig: running-retrain} we can see that AdaCompress can adaptively update the estimation probability $ p_{\rm est} $, usually, when the overall accuracy decreases, AdaCompress will increase the estimation probability, trying to catch the scenery change. When the overall accuracy is stable and high enough, the estimation probability $ p_{\rm est} $ decreases to reduce transmission. 

Upon the data scenery change shown as the vertical red line in Figure \ref{fig: running-retrain}, comparing to the earlier steps, the accuracy decreases dramatically and therefore $ p_{\rm est} $ raises to determine whether scenery changes, the accuracy keeps dropping in the following estimations. Therefore, AdaCompress starts to retrain, to adapt the RL agent into the current scenery. The retrain steps are shown as the light-blue region in Figure ~\ref{fig: running-retrain}. In the retrain phase, AdaCompress always uses the reference image's prediction label $ \vec{y}_{\rm ref} $ as the output result, therefore the accuracy $ \mathcal{A} $ and $ p_{\rm est} $ is locked to 1. After finishing retraining the agent in the new scenery, in the following iterations, sometimes the accuracy decrease accidentally, the estimation probability $ p_{\rm est} $ also raises to get more samples, but the accuracy is not lower than the initial average accuracy $ \mathcal{A}_0 $ of this scenery, therefore the retrain phase will not be triggered again. 

From Figure \ref{fig: running-retrain} we can also observe the uploading file size overhead in different phases, we can see that in retrain phase, AdaCompress uploads more data than the conventional benchmark, but in inference phase, AdaCompress's upload data size is only half of the benchmark's. 

\subsection{End-to-end latency simulation}

Comparing to the conventional solution that uploads the image directly, in our solution, the image is passed to the DRL agent first to estimate the compression level. Running this DRL agent brings extra latency to the whole system. In this subsection, we evaluate this latency overhead. 

We tested the DRL agent's inference time and compressed file size for batches of images, and simulate the latency of uploading such compressed images. We test the average inference latency from 1000 ImageNet images and simulate the network bandwidth as 27.64 Mbps according to the global average fixed broadband upload speed~\cite{speedtest} in Feb. 2019 to verify the end-to-end latency performance. The latency comparison is listed in Table \ref{tab: latency-overhead}. 

\begin{table}[htbp]
	\begin{tabular}{lll}
		\toprule
		& Benchmark & AdaCompress  \\ \midrule
		Average upload size          & 42.68 KB  &  18.46 KB            \\
		Inference latency    & 0 s       & 2.09 ms          \\
		Transmission latency & 12.35 ms       & 5.34 ms          \\
		Overall latency      & 12.35 ms       & 7.43 ms          \\ \bottomrule
	\end{tabular}
	\caption{Latency between image upload and inference result feedback}
	\label{tab: latency-overhead}
	\vspace{-0.5cm}
\end{table}

Our solution brings in inference latency to the end-to-end latency, but the transmission latency is much lower by shrinking the upload file size. In today's network architecture where the edge infrastructure's computational power is increasing significantly ~\cite{satyanarayanan2017emergence,hu2015mobile}, we can use the computing power of the edge infrastructure in exchange for the reduction of upload traffic and transmission latency. 

\section{Related Works}
\label{sec: related_works}

As cloud-based computer vision services have become the norm for today's applications ~\cite{huynh2017deepmon, agrawal2015cloudcv}, many studies have been devoted to improving the cloud-based model execution, including model compression and data compression.

\subsection{Model compression}

Though the accurate term is still for the community to debate, we use ``model compression'' to represent the studies on {\em compressing} and {\em moving} the deep learning models close to users. A number of studies tried to compress the deep learning models and deploy them \emph{locally} ~\cite{prun_quanti, pruning_han, quantize, quantize_3bit, quantization, structured_pruning}, i.e., running an alternative ``smaller version'' of a computer vision model at the user end, to avoid the image upload, so that to improve the inference efficiency. Other studies proposed to run part of a deep learning model locally ~\cite{ILP_Decoupling, jalad, Edge_LBP, Neurosurgeon}, by decoupling the deep learning model into different parts, e.g., based on the layers in the deep learning model, so that a part of the inference is done locally to save some execution time. However, these solutions usually need to \emph{re-train} the model, using the original dataset of the model, which is not practical for today's cloud computer vision services that are merely a black box to end-users, e.g., in the form of a RESTful API.

\subsection{Data compression}

Data compression solutions study how to compress the original data (e.g., a video or image) to be inferred by the cloud deep learning model, so that less traffic is used to upload the data to improve inference speed. Conventional data compression solutions (e.g., JPEG, WebP, JPEG2000 etc.) and some recent neural network based compression solutions ~\cite{toderici2017full,theis2017lossy,toderici2015variable, rippel2017real} are initially designed for human vision systems. In recent years, researchers start to found that the human visually optimized data compression solutions are not usually applicable to deep learning vision systems. Delac et al.~\cite{delac2005effects} observed that, in some cases, higher compression level does not always deteriorate the model inference accuracy, and in some cases, even improves it slightly. Dodge et al.~\cite{dodge2016understanding} further discovered that besides the JPEG compression, four types of quality distortions: blur, noise, contrast, and JPEG2000 compression can also affect the performance in deep learning inference. 

Based on these insights, Robert et al. ~\cite{torfason2018towards} tried to train the neural network from the compressed representations of an auto-encoder. Liu et al.~\cite{DeepN-JPEG} proposed DeepN-JPEG that provides a JPEG quantization table learned from the dataset so that the compressed image size is reduced for deep learning models. Recently, Lionel et al. ~\cite{gueguen2018faster} present a new type of neural network that inference directly from the discrete cosine-transform (DCT) coefficients in the middle of the JPEG codec. Baluja et al.~\cite{baluja2019task} proposed task-specific compression that compresses images based on the end-use of the image. 

However, such proposals all need one to understand the characteristics of the cloud-end deep learning model and have access to the original training dataset, to generate the appropriate color space and/or compression schemes. To the best of our knowledge, we are the first to propose an adaptive compression configuration solution that learns the deep learning model by itself. 

\section{Conclusion and future work}
\label{sec: conclusion}

To reduce the upload traffic load of deep learning application, most researchers focus on modifying the deep learning model, but this does not apply to the industry because the backend deep model is usually inaccessible for users. We present a heuristic solution using a deep learning agent to decide the proper compression quality for each image, according to the input image and backend service. Our experiments show that for different backend deep learning cloud services and different input image scenery, using different quality selection strategy can significantly reduce the upload file size while keeping comparable accuracy. Based on this work, some possible future orientations can focus on the following: 1) In some regularly change scenery (e.g., daytime and nighttime, etc.), one can design an agent caching strategy, to cache an agent for a specific scenery and use it again when a similar scenery arrives rather than retrain from scratch. 2) By introducing transfer learning and knowledge distillation, an agent could learn from another nearby agent to accelerate its training.

%
\begin{acks}
This work is supported in part by NSFC under Grant 61872215, 61531006, 61771273 and U1611461, National Key R\&D Program of China under Grant 2018YFB1800204 and 2015CB352300, SZSTI under Grant JCYJ20180306174057899 and JCYJ20180508152204044, and Shenzhen Nanshan District Ling-Hang Team under Grant LHTD20170005.
\end{acks}

\bibliographystyle{ACM-Reference-Format}
\balance
\bibliography{main-base}

\end{document}